%%%%%%%%%%%%%%%%%%%%%%%%%%%%%%%%%%%%%%%%%%%%%%%%%%%%%%%%%%%%%%%%%%%%%%
% STEADY STATES OF HARMONIC OSCILLATOR CHAINS
% AND SHORTCOMINGS OF HARMONIC HEAT BATHS
% Published in Physica A, 202, 342 (1994)
% By Max Tegmark and Leehwa Yeh
% Email: max@mppmu.mpg.de
% http://www.mpa-garching.mpg.de/~max/steady.html   (faster from Europe)
% http://astro.berkeley.edu/~max/steady.html        (faster from the US)
%%%%%%%%%%%%%%%%%%%%%%%%%%%%%%%%%%%%%%%%%%%%%%%%%%%%%%%%%%%%%%%%%%%%%%
% This is a plain TeX file (not LATeX)
% :-)
%%%%%%%%%%%%%%%%%%%%%%%%%%%%%%%%%%%%%%%%%%%%%%%%%%%%%%%%%%%%%%%%%%%%%%
%
% EQUATION NUMBERING STUFF:
%
\def\inc#1{\hbox{\global \advance#1 1}}
\countdef\eqnr=1    
\eqnr=1           
\def\nexteq{\inc{\eqnr}}    
\def\enr{\number\eqnr\nexteq}   
\def\eq{\eqno(\enr)}   

\countdef\HamEq=21
\countdef\EvolEq=22
\countdef\CorrEq=23
\countdef\EnergyEq=24
\countdef\WeakCorrEq=25
\countdef\DefEq=26
\countdef\SuddenUeq=27
\countdef\fmnEq=28
\countdef\fmnEqTwo=29
\countdef\Ueq=30
\countdef\SteadyEq=31
\countdef\SuddenDeq=32
\countdef\OneModeAdiabEq=33
\countdef\AdiabUeq=34
\countdef\AdiabDeq=35
\countdef\Uodeeq=36
\countdef\FKMeq=37
\countdef\InitialSSeq=38

%
% REFERENCE STUFF:
%
\countdef\refnr=2
\refnr=1

\countdef\Huerta=3
\countdef\FKM=4
\countdef\MMref=5
\countdef\FLO=6
\def\nextref{\inc{\refnr}} 
\def\rnr{$\number\refnr$\nextref} 
\def\ref{[\rnr]}
\def\refn{\par\pp{\bf \rnr.} }
\def\pp{\parshape 2 0truecm 17truecm 2truecm 15truecm}
% This gives references Phys. Rev. A style:
\def\rf#1;#2;#3;#4;#5 {\par\pp{\bf \rnr.} #1, {#2}, 
{\bf #3}, #4 (#5)\par}

\def\ie{{\it i.e.}}
\def\tr{\hbox{tr}\>}

\def\header#1{\bigskip\goodbreak{\bf#1}\smallskip}
\def\llist{\noindent\parshape 2 0.5cm 15.0cm 1.32cm 14.18cm}
\def\lllist{\noindent\parshape 2 1.5cm 14.0cm 2.32cm 13.18cm}
\def\slist{\noindent\parshape 2 0.5cm 15.0cm 1.00cm 14.50cm}

\def\arcoth{\coth^{-1}}

\def\om{\omega_0}
\def\vm{{\bf\mu}}
\def\vz{{\bf z}} \def\vp{{\bf p}} \def\vq{{\bf q}}
\def\oz{{\bf\hat z}} \def\op{{\bf\hat p}} \def\oq{{\bf\hat q}}
\def\opH{{\hat H}}
\def\zzt{\vz\vz^T} \def\mmt{\vm\vm^T}
\def\diag#1{\>{\rm diag}\{#1\}}
\def\expec#1{\langle#1\rangle}
\def\oh{{1\over 2}}
\def\eot{{\epsilon\over 2}}
\def\Aoh{A^{1/2}}  \def\Amoh{A^{-1/2}}
\def\abrac{\left[\Amoh-\Aoh\right]}

\def\hootktz{{\hbar\om\over 2kT_0}}
\def\hoaotkt{\beta\Aoh}
\def\cth{\coth\hoaotkt}
\def\Aot{\left[\Aoh t\right]}
\def\tAot{\left[2\Aoh t\right]}

\def\Cstat{C_1}  \def\Cosc{C_2}
\def\ootp{{1\over 2\pi}}
\def\ipp{\int_{-\pi}^{\pi}}

\def\imii{\int_{-\infty}^{\infty}}
\def\iww{\int_{\omega_{min}}^{\omega_{max}}}
\def\lth{\lambda^2(\theta)}
\def\sumk{\sum_{k=-N}^{N}}
\def\suml{\sum_{l=-N}^{N}}
\def\uh{\hat u}
\def\gh{\hat g}
\def\Th{\Theta(t)}
\def\sinth{\sin\theta(t)}  \def\costh{\cos\theta(t)}
\def\sinTh{\sin\Th}  \def\cosTh{\cos\Th}

\def\Aoht{A^{1/2} t}

\def\sinhmess{a}
\def\crr{\cr\noalign{\vskip 4pt}}

\baselineskip12pt
\hoffset0.5cm
\hsize 15.5cm

% JUNK FOR NICE COVER PAGE

\font\titlefont=cmb10 at 15truept
\font\namefont=cmr12 at 14truept
\font\addrfont=cmti12
\font\rmtwelve=cmr12

% ABSTRACTS:
\newbox\abstr
\def\abstract#1{\setbox\abstr=
    \vbox{\hsize 5.0truein{\par#1}}
    \centerline{ABSTRACT} \vskip12pt 
    \hbox to \hsize{\hfill\box\abstr\hfill}}

%  macro for invoking today's date (when TeX is run on your file)
\def\today{\ifcase\month\or
        January\or February\or March\or April\or May\or June\or
        July\or August\or September\or October\or November\or December\fi
        \space\number\day, \number\year}

\def\author#1{{\namefont\centerline{#1}}}
\def\addr#1{{\addrfont\centerline{#1}}}

{ % SETTINGS FOR COVER PAGE ONLY
\rmtwelve

%\hoffset=1 truein
%\voffset=1 truein
\vsize=9 truein
\hsize=6.5 truein
%\parskip=0.2truecm
\raggedbottom
\baselineskip=16pt

July 13, 1993
\vskip1.2truecm
\centerline
{\titlefont STEADY STATES OF HARMONIC OSCILLATOR CHAINS}
{\titlefont AND SHORTCOMINGS OF HARMONIC HEAT BATHS
\footnote{$^\dagger$}{\rm Published in Physica A, {\bf 202}, 342 (1994)}} 
\nobreak
  \vskip 1.2truecm
 
  \author{Max Tegmark}
  \smallskip
  \centerline{and}
  \smallskip
  \author{Leehwa Yeh}
  \bigskip
  \addr{Department of Physics, University
of California,}
  \addr{Berkeley, California  94720}
  \addr{Email: max@physics.berkeley.edu, yehl@csa.lbl.gov}
  \bigskip

  \vskip 0.8truecm
 
\abstract{
We study properties of steady states (states with
time-independent density operators) of systems of coupled
harmonic oscillators. Formulas are derived showing how adiabatic
change of the Hamiltonian transforms one steady state into
another. It is shown that for infinite systems, sudden change of
the Hamiltonian also tends to produce steady states, after a
transition period of oscillations.
These naturally arising steady states are compared to the
maximum-entropy state (the thermal state) and are seen 
not to coincide in general. 
The approach to equilibrium of subsystems consisting of $n$
coupled harmonic oscillators has been widely studied, but 
only in the simple case where $n=1$. The power of our results is
that they can be applied to more complex subsustems, where
$n>1$.  It is shown that the use of coupled harmonic
oscillators as heat baths models is fraught with some
problems that do not appear in the simple $n=1$ case.
Specifically, the thermal states that are though to be
achievable through hard-sphere collisions with heat-bath
particles can generally {\it not} be achieved with harmonic
coupling to the heat-bath particles, except approximately when
the coupling is weak.}

\bigskip\bigskip
\centerline{PACS codes: 05.70.Lu, 05.30.Ch, 03.65.Ge} 
 
}  % End of cover settings 
\vfill\eject
 
%
% HERE COMES THE MAIN PART OF THE PAPER:
% 

\baselineskip12pt
\hoffset0.5cm
\hsize 15.5cm

\beginsection{I. INTRODUCTION}

Quantum systems with quadratic
Hamiltonians have been studied extensively. One reason for this
is that they are just about the only quantum systems whose
time evolution can be found analytically. Hence they have
provided useful and tractable models in a wide range of fields,
including statistical mechanics, quantum optics and solid state
physics.

In this paper, we derive some useful formulas for the
subset of such systems known as harmonic chains. These may
be pictured as a chain of coupled harmonic oscillators where
the coupling between any two oscillators depends only on the
distance between them.
We derive the most general steady states of coupled harmonic
oscillators, and then study how adiabatic and sudden changes of
the Hamiltonian convert one such steady state into another.

The main motivation for studying harmonic
chains has been the pursuit of a dynamical basis for
equilibrium statistical mechanics. 
In short, we frequently assume that the density matrix of
a system is $\rho\propto e^{-H/kT}$, {\ie} the
thermal state corresponding to a temperature $T$, and it would
be rather embarrassing if we had no completely solvable
heat bath model that explicitly evolved systems into such states.
An excellent summary of
the early developments is given by Huerta and Robertson
(see 
\Huerta=\refnr
{\ref}
and references therein). 
The main idea is to study the time evolution of the mean
values and correlations of some small subset of the oscillators,
called the {\it system}, by taking a
partial trace over the rest of the oscillators, called
the {\it heat bath}. The goal is
to show that the system will exhibit standard thermodynamic
features such as Brownian motion and approach to thermal
equilibrium.

Numerous models have been developed that show
such features, and we will give only a brief overview here.
The dynamics of an
infinite, one-dimensional chain of oscillators bound only to
their nearest neighbors was solved in 1914 by Schr\"odinger
{\ref}. Klein and Prigogine {\ref} used these results to show
that suitable initial conditions for the heat bath led to
equipartition of energy in the system. Hemmer {\ref} studied the
same model in more detail and found that the heat bath could
produce Brownian motion of the system. Results for more general
quadratic Hamiltonians have since been derived by numerous
authors \FKM=\refnr
$[\number\refnr-
\nextref
\FKM=\refnr
\nextref\nextref\nextref\nextref\nextref
\number\refnr]$. 
\nextref
An up-to-date overview of the
models is given by Ford, Lewis and O'Connell 
\FLO=\refnr
(see {\ref} and
references therein), who summarize six successful models and
show that they are all unitarily equivalent.
Most of the recent literature on the subject has focused on
reduced descriptions of a single oscillator in the chain
such as a Langevin or Fokker-Planck equation.

Thus the approach to equilibrium of subsystems consisting of $n$
coupled harmonic oscillators has been widely studied, but 
only in the special case where $n=1$. 
This is rather unfortunate, since as we shall see, some
interesting complications can arise when $n>1$.
We will study subsystems consisting of an arbitrary number of
oscillators, including the seldom mentioned case where the
mean values of the heat bath oscillators are non-zero. 
We do this using a model involving cyclic matrices, similar to
that of $[\number\FKM]$. As will be seen below, the great
advantage of this model is that subsystems with large
$n$ can be treated with the same ease as the $n=1$ case, since
we need not distinguish beetween the system and the heat bath
{\it a priori}.

This paper is organized as follows:
In section II, we establish some notation and derive the most
general steady state of coupled harmonic oscillators. In
section III, we review some useful properties of cyclic matrices.
In sections IV, V and VI, we study how adiabatic and sudden
changes of the Hamiltonian transform one steady state into
another. Finally, in sections VII and VIII we
compare these naturally arising steady states with thermal
states and discuss some shortcomings of coupled harmonic
oscillators as heat bath models. 
A rigorous convergence proof for the sudden change case is
given in Appendix A.

\beginsection{II. THE GENERAL STEADY STATE}

In this section, we establish some notation and derive the
covariance matrix of the most general time-independent state of
coupled harmonic oscillators of equal mass.

As our quantum system, let us take $2N+1$ coupled harmonic
oscillators of equal mass, labelled 
${-N,...,-1,0,1,...,N}$.
Denoting a point in
the $2(2N+1)$-dimensional phase space by 
$$\vz = \pmatrix{\vq\cr\vp}\eq$$ 
and the corresponding operators by
$$\oz = \pmatrix{\oq\cr\op},\eq$$
we can write the Hamiltonian as
\HamEq=\eqnr
$$\opH = {1\over 2m}\op^T\op + 
{m\om^2\over 2}\oq^T A\oq,\eq$$ 
where the time-dependent matrix $A$ is symmetric and positive
definite. Throughout this paper, we will use units
where $m=\om=1$. The number of oscillators can be
either finite or infinite.
Let us define the mean vector $\vm$ and the covariance matrix $C$
by  
$$\cases{
\vm&$\equiv\expec{\oz}$,\cr
\noalign{\vskip 4pt}
C&$\equiv 
\pmatrix{
\expec{\oq\oq^T}&\oh\expec{\oq\op^T+\op\oq^T}\cr
&\cr
\oh\expec{\oq\op^T+\op\oq^T}&\expec{\op\op^T}
}
 - \vm\vm^T$.}\eq$$
(The symmetric ordering is necessary since $\oq$ and $\op$ do not
commute.) It is well-known that since $\opH$ is quadratic, the
time evolution of $\vm$ and $C$ is identical in classical and
quantum statistical mechanics and is given by \EvolEq=\eqnr
$$\cases{
\vm(t)& = $U(t)\vm(0)$,\cr
\noalign{\vskip 4pt}
C(t)& = $U(t)C(0)U(t)^T$,\cr
}\eq$$
where the time-evolution matrix is defined by
\Uodeeq=\eqnr
$$\cases{
{d\over dt}U = \pmatrix{0&I\cr -A(t)&0} U,\crr
U(0) = I.
}\eq$$
In the special case where the matrix $A$ is independent of $t$,
this equation can be integrated, yielding 
\DefEq=\eqnr
\footnote
{$^\dagger$}
{
Here and throughout this paper, the action of a function on a
symmetric matrix is  defined as the corresponding real-valued
function acting on its eigenvalues:
Since all symmetric matrices $A$ can be diagonalized as
$$A=R\Lambda R^T,$$
where $R$ is orthogonal and $\Lambda = \diag{d_i}$ 
is diagonal and real, we can extend any mapping $f$
on the real line to symmetric matrices by defining
$$f(R\diag{d_i}R^T) \equiv
R\diag{f(d_i)}R^T.\eq$$
It is easy to see that this definition is consistent with
power series expansions whenever the latter converge.
For example, 
$$\cos\Aoh = \sum_{n=0}^{\infty}{(-1)^n\over(2n)!}A^n.$$
} 
\nobreak
\SuddenUeq=\eqnr
\Ueq=\eqnr
$$U(t)=\exp\left[\int_0^t\pmatrix{0&I\cr
-A&0}\,d\tau\right]
= \pmatrix{\cos\Aoht&A^{-1/2}\sin\Aoht\cr 
-A^{1/2}\sin\Aoht&\cos\Aoht}.\eq$$

One useful feature of 
$\mu$ and $C$ 
is that if the
phase-space distribution (the Wigner distribution {\ref}) is
Gaussian, then they determine the state completely. For all the
steady states considered below, this will indeed be the case.
(Such Gaussian states in quantum optics are often
referred to as multimode thermal squeezed coherent
states, but we will refer to them as Gaussian for
brevity.)
For a good review of Wigner distributions, see {\ref}, {\ref}
and references therein.
 
$\mu$ and $C$ are also very useful for studying subsystems, 
as we will do in sections VII and VIII.
Rather than having to take cumbersome partial traces
of the density operator, the means and internal 
correlations of any subsystem can simply be read off as 
subsets of the entries in $\mu$ and $C$. If the state of
the total system is Gaussian, so is the state of the subsystem,
which means that the latter is uniquely determined by these
subsets.

By a {\it steady state} of a time-independent Hamiltonian, we
mean a state whose density operator 
(or equivalently its Wigner function)
is independent of time. 
The most general steady state is clearly 
a statistical mixture of energy eigenstates, {\ie}
a density operator of the form
$$\hat{\rho} = \sum w_n |E_n\rangle\langle E_n|,\eq$$
where $w_n \geq 0$ and $\sum w_n = 1$. In terms of the Wigner
function, this becomes  
$$W(\vq,\vp) = \sum w_n W_n(\vq,\vp).\eq$$ 
Since the energy eigenstates $W_n$ are cumbersome
products of Laguerre polynomials, this expression does not
shed much light on the physical nature of steady states. 
Fortunately, we can obtain a simple and useful result for their
covariance matrices:
By differentiating Eq. $(\number\EvolEq)$ with respect to
$t$ using Eq. $(\number\SuddenUeq)$, it is readily shown that
the mean vector and the covariance matrix will remain constant
over time if and only if they are of the form 
\SteadyEq=\eqnr
$$\cases{
\vm& = $0$,\crr
C& = $\pmatrix{D&0\cr 0&AD}$,}\eq$$
where $D$ is a symmetric, positive definite matrix that commutes
with $A$, and $A$ is independent of time. 
We will discuss the case $\vm\neq 0$ in Section VIII.
Note, however, that a covariance matrix of the above form will
remain constant over time regardless of the value of $\mu$.

In section IV, we will see how adiabatic change transforms
one such steady state into another.
In section V, we will see that these steady states are a
form of attractors, in that quite general states tend to
converge towards them as $t\to\infty$.
In section VI, we will study a one-parameter subset of
steady states, the thermal states, and compare them with
the states arising for adiabatic and sudden change.

\beginsection{III. CYCLIC MATRICES}

In many parts of this paper, we assume that the
potential matrix $A$ is {\it cyclic}, which means that we can
write $A_{ij} = a_{|i-j|}$ and interpret the system as a chain of
harmonic oscillators where the coupling between any two
oscillators depends only on the separation between them. For
finite $N$,  we will for computational simplicity identify $N+i$
with $-N-1+i$ and picture a ring  of oscillators rather 
than an array with two ends (thus $A_{ij}=A_{kl}$
if $i-j=k-l$ mod $(2N+1)$).
Using $(\number\DefEq)$, we can write any function of a
(cyclic or non-cyclic) matrix $A$ as 
\fmnEq=\eqnr
$$f(A)_{mn} = \sumk R_{mk}R_{nk}f(\lambda^2_k).\eq$$

Cyclic matrices have the great advantage that they all commute.
This is because 
they can all be diagonalized by the same matrix $R$, an
orthogonal version of the discrete Fourier matrix. 
Physically, this means that plane waves form a complete set of
solutions.
Ford, Kac and Mazur $[\number\FKM]$ (hereafter referred to as
FKM) show that if $A$ is symmetric, positive-definite, cyclic and
infinite-dimensional, then Eq. $(\number\fmnEq)$ reduces to 
\FKMeq=\eqnr
$$f(A)_{mn} = \ootp\ipp d\theta
f\left[\lth\right]
\cos(m-n)\theta,\eq$$ where the
spectral function $\lth$ is 
the function whose Fourier coefficients are row zero of $A$.
The spectral function can be interpreted as
a dispersion relationship, $\lambda$ being the frequency of a
wave with wave number $\theta$. 
Note that $f(A)$ is cyclic as well, {\ie} its
components depend only on the distance to the diagonal.

A cyclic potential frequently discussed in the literature is
the {\it nearest neighbor potential}, the 
case where each mass is coupled only to a fixed spring and to
its nearest neighbor: 
$$\opH = \sumk\left[{1\over 2}\hat p_k^2 + 
{1\over 2} \hat q_k^2 +
{\gamma^2\over 2}
\left(\hat q_{i+1}-\hat q_i\right)^2\right],\eq$$  
{\ie} $A_{kk} =
1+2\gamma^2$, $A_{k,k\pm1} = -\gamma^2$ and all other elements
of $A$ vanish. For this special case, the spectral function is  
$$\lambda^2(\theta) = 1 + 4\gamma^2\sin^2{\theta\over
2}.\eq$$

\beginsection{IV. ADIABATIC CHANGE}

In this section we will study the time evolution of 
$C$ during an adiabatic change of the Hamiltonian of a steady
state.

For generic time-dependent cyclic Hamiltonians, the
time-evolution operator $U(t)$ depends on the entire history
$A(\tau)$, ${0\leq\tau\leq t}$, according to Eq.
$(\number\Uodeeq)$.  For the case where $A$ changes extremely
slowly relative to the eigenfrequencies, this simplifies so that
$U(t)$ depends only on the values of $A$ at $\tau=0$ and
$\tau=t$, and is independent of $A$ at 
intermediate times.
According to a well-known result, the
time evolution of a single harmonic oscillator $(N=0)$ where the
frequency $\omega$ changes adiabatically is given by {\ref}
\OneModeAdiabEq=\eqnr 
$$U(t) = 
\pmatrix{\omega(t)^{-1/2}&0\cr 0&\omega(t)^{1/2}}
\pmatrix{\costh&\sinth\cr-\sinth&\costh}
\pmatrix{\omega(0)^{1/2}&0\cr 0&\omega(0)^{-1/2}},
\eq$$
where
$$\theta(t)\equiv\int_0^t \omega(\tau) d\tau.\eq$$
We are interested in the case of arbitrary $N$.
Since we are assuming that the matrices $A(t)$ are cyclic
for all $t$, they can all be diagonalized by the same
time-independent matrix $R$,
$$A(t) = R^T \diag{\omega_i(t)^2} R,$$
where the eigenfrequencies $\omega_i$ also change adiabatically.
Hence we can apply
$(\number\OneModeAdiabEq)$ to each mode separately, which after
employing $(\number\DefEq)$ yields
\AdiabUeq=\eqnr
$$U(t) = 
\pmatrix{A(t)^{-1/4}&0\cr 0&A(t)^{1/4}}
\pmatrix{\cosTh&\sinTh\cr-\sinTh&\cosTh}
\pmatrix{A(0)^{1/4}&0\cr 0&A(0)^{-1/4}},\eq$$
where
$$\Theta(t)\equiv\int_0^t A(\tau)^{1/2}   d\tau.\eq$$
If our system is in a cyclic steady state initially, {\ie}
if $C(0)$ is cyclic, then $D(0)$ is
cyclic and commutes with all the above matrices, so  substituting
$(\number\AdiabUeq)$ into $(\number\EvolEq)$ yields
$$C(t) = \pmatrix{D(t)&0\cr 0&A(t)D(t)},\eq$$
where
$$D(t) = \left[A(0)A(t)^{-1}\right]^{1/2}D(0).\eq$$
Note that the quantity $A(t)D(t)^2$ is an 
{\it adiabatic invariant}, {\ie} stays constant over time. We
see that adiabatic change forms an Abelian transformation group
on the set of all steady states, and that any given steady state
can be transformed into exactly those steady states that have
the same adiabatic invariant.

\beginsection{V. SUDDEN CHANGE FOR INFINITE SYSTEMS}

In this section we will study the time evolution of $C$ after a
sudden change in the Hamiltonian of a steady state.
We will
mainly be interested in the following question: Does $C(t)$ keep
oscillating forever, or does the state converge towards a new
steady state as $t\to\infty$?

If the system is in a steady state
\InitialSSeq=\eqnr
$$C(t) = \pmatrix{D_0&0\cr 0&A_0 D_0}\eq$$
for $t<0$ and the potential matrix in the Hamiltonian
$(\number\HamEq)$ changes abruptly from $A_0$ to a new constant
value $A$ at $t=0$, then the time evolution for $t>0$ will be
given by $(\number\EvolEq)$ and 
$(\number\SuddenUeq)$.
Let us temporarily choose the simple initial conditions
$A_0=D_0=I$, and return to the general case later.
For this choice, $C(0)=I$, so that Eq. $(\number\EvolEq)$
gives $C(t) = U(t)U(t)^T$, \ie
$$C(t) = 
\pmatrix{
\cos^2\Aot + A^{-1}\sin^2\Aot&
\abrac \sin\Aot\cos\Aot\cr
\abrac \sin\Aot\cos\Aot&
\cos^2\Aot + A\sin^2\Aot}.\eq$$
We can separate this $C(t)$ into a time-independent part and an
oscillating part with zero time average by writing  
$C(t) = \Cstat + \Cosc(t)$, where 
$$\cases{
\Cstat& = ${1\over 2}\pmatrix{
I+A^{-1}&0\cr
0&I+A}$,\cr
\noalign{\vskip 6pt}
\Cosc(t)& = $ 
{1\over 2}\pmatrix{
\left[I - A^{-1}\right]\cos\tAot&
\abrac \sin\tAot\cr
\abrac \sin\tAot&
\left[I - A\right]\cos\tAot}.$\cr
}\eq$$
In Appendix A, we show that for a generic cyclic
infinite-dimensional symmetric positive definite matrix $A$, 
$\Cosc(t)\to 0$ as $t\to\infty$, so that $C(t)$ roughly speaking
evolves  as follows: 
$$\pmatrix{I&0\cr0&I} 
\to
\pmatrix{
{\rm Time-dep.}\cr
{\rm mess}}
\to
\pmatrix{
{1+A^{-1}\over 2}&0\cr
0&{1+A\over 2}}
\quad\hbox{as }t\to\infty.\eq$$
These results are readily generalized to arbitrary cyclic
initial conditions at $t=0$:
$$\pmatrix{X&Z\cr Z&Y} 
\to
\pmatrix{D&0\cr 0&A D}
\quad\hbox{as }t\to\infty,\eq$$
where
$$D = \oh\left[X+ A^{-1}Y\right].\eq$$
If the initial state is the steady state
$(\number\InitialSSeq)$, this reduces to 
\SuddenDeq=\eqnr
$$D = \oh\left[I+ A_0 A^{-1}\right] D_0.\eq$$
The fact that all matrices are cyclic is crucial, since it means
that they can all be simultaneously diagonalized and hence all
commute.

Note that whereas adiabatic change formed a transformation
group on the set of steady states, sudden change (and
subsequently waiting until the oscillations have died down) does
not. The steady state reached through a sequence of two sudden
changes, one after the other, can in general not be achieved
with a single sudden change. This is obvious from the
fact that no more than half of the energy can be lost in single
sudden change.

A second difference from the adiabatic case is that sudden
changes in general have no inverse, {\ie} cannot be undone.
If the potential is suddenly changed from $A_0$ to $A$ and then,
after the oscillations have died down, back to $A_0$, the net
result is 
$$D\mapsto {1\over 4}\left[2I+A_0^{-1}A + A_0 A^{-1}\right]D.\eq$$
This steady state has a greater energy than the initial one
for any $A\neq A_0$.

\beginsection{VI. SUDDEN CHANGE FOR LARGE BUT FINITE SYSTEMS}

The result $\Cosc(t)\to 0$ as $t\to\infty$ does not hold for
finite $N$. Indeed, for finite $N$, the components of $\Cosc(t)$
must return to their initial values an infinite
number of times, the typical time between such recurrences
being the Poincar\'e recurrence time.  
How can this be reconciled with the results of the previous
section? An excellent discussion of these matters is given by
Mazur and Montroll $[6]$ for some special cases.
Here we will limit ourselves to an informal heuristic
discussion, whose aim is to give a qualitative feeling for the
non-cyclic case and the
$N\to\infty$ limit.

Applying Eq. $(\number\fmnEq)$ to the components of the
various terms in $\Cosc(t)$, the function $f(\lambda^2)$ always
contains a trigonometric factor that oscillates increasingly
rapidly with respect to $\lambda^2$ as $t\to \infty$.
Suppose for a moment that the product $R_{mk}R_{nk}$ varies
smoothly with $k$ and that the spectrum $\lambda_k^2$ is not too
degenerate.  Then for very large $t$, all phase coherence
between different terms in the sum is lost and we can for all
practical purposes replace the arguments of the trigonometric 
functions by random variables $\phi_i$ that are uniformly
distributed on the interval $[0,2\pi]$. 
Using $\expec{\cos\phi_i} = 0$ and 
$\expec{\cos\phi_i\cos\phi_k} = \oh\delta_{ik}$, this means
that if we let $t$ be a random variable uniformly
distributed on the interval $[-T,T]$ for some large enough
$T$, then the expectation value and variance of a
typical term like  
$$X(t)\equiv A\cos\tAot\eq$$ 
is approximately given by 
$$\expec{X(t)_{mn}} = 
\expec{\sumk R_{mk}R_{nk}\lambda_k^2
\cos\left(2\lambda_k t\right)} =
\sumk R_{mk}R_{nk}\lambda_k^2\expec{\cos\phi_k} = 0,\eq$$

$$\expec{X(t)_{mn}^2} =
\sumk\suml
R_{mk}R_{nk}R_{ml}R_{nl}\lambda_k^2\lambda_l^2
\expec{\cos\phi_k\cos\phi_l}
= \oh\sumk \left(R_{mk}R_{nk}\lambda_k^2\right)^2.\eq$$
Since the typical element of a generic $N\times N$ orthogonal
matrix $R$ is of order $N^{-1/2}$, the sum above will be of
order $N\times N^{-2}$, which yields a standard deviation of
order $N^{-1/2}$. 
Thus for large but finite systems, we expect the elements of
$\Cosc(t)$ to evolve as follows:

\llist{
(i) During an initial transition period whose duration
is of the order of the dynamical time scale $\omega_0^{-1}$, they
decay from their initial values of order unity to values of order
$N^{-1/2}$.}

\llist{
(ii) After that, they oscillate around zero with an
oscillation amplitude of order $N^{-1/2}$.}

\llist{
(iii) Since their time evolution is almost periodic, they must
return to values of order unity an infinite number of times.
This happens approximately once every Poincar\'e recurrence
time.
However, as shown by Mazur and Montroll, the Poincar\'e 
time scale is generally enormous compared to the dynamical
time scale, since it tends to grow exponentially with $N$ for
systems of this type.}

\beginsection{VII. IMPLICATIONS FOR HEAT BATH MODELS}

In this section we will discuss thermal states, and compare them
to the stationary states arising from adiabatic and sudden
change.
It will be seen that the latter two are in general not thermal.

\header{Thermal states}

The energy of a steady state is 
$$E=\expec{H} = 
{1\over 2}\tr\pmatrix{A&0\cr 0&I}\pmatrix{D&0\cr 0&AD} = 
\tr AD.\eq$$
Given a Hamiltonian and a fixed energy, the 
{\it thermal state} is the 
state that has the maximum entropy consistent with this energy.
The Wigner 
function of a thermal state is Gaussian. FKM show that a
thermal state of temperature $T$ is the steady state
given by 
$$\cases{
\vm& = $0$\cr
C&= $\oh\pmatrix{A^{-1/2}\cth&0\cr 0&{A^{1/2}\cth}}$
},\eq$$
where we have defined $\beta\equiv {\hbar\om\over 2kT}$.
In the classical limit $\beta\ll 1$, this reduces
to 
$$C = {1\over 2\beta}\pmatrix{A^{-1}&0\cr 0&I},\eq$$
whereas $T=0$ yields the ground state
$$C = \oh\pmatrix{A^{-1/2}&0\cr 0&A^{1/2}}.\eq$$

\header{Heat bath models}

According to the standard interpretation,  a system will be
in a thermal state if it has been exposed to an ideal heat bath 
long enough to reach thermal equilibrium.
(After this the heat bath can be removed if the system is kept
isolated, since the thermal state is a steady state.) 
Numerous attempts have been made to model ideal heat
baths by infinite systems of coupled harmonic oscillators. We
will now use the results we have derived for adiabatic and sudden
change to illustrate some shortcomings of such models, and
clarify what they can and can not do. 

Typically, the time evolution of some small subset of
oscillators, called the {\it system}, is studied by taking a
partial trace over ({\ie} ignoring) the rest of the oscillators,
called the {\it heat bath}. A goal of all such pursuits
has been to devise interaction Hamiltonians $A(t)$ such that
the state of the subsystem becomes thermal.

Our use of $\mu$ and $C$
has the advantage that we need
not distinguish between system and heat bath {\it a priori}: if
we decide to view some subset of oscillators as the system,
then their means and internal correlations are simply
given as subsets of the entries in $\mu$ and $C$.

We can summarize our previous results for steady states as
follows: 
$$\cases{
D\mapsto \left(A_0A^{-1}\right)^{1/2} D&in adiabatic case,\crr
D\mapsto \oh\left[I + A_0A^{-1}\right] D&in sudden case,\crr
D= \oh\Amoh\cth&in thermal case.
}\eq$$
From this it readily follows that
we can transform an arbitrary steady state given by $A_0$ and
$D_0$ into a thermal state with an arbitrary temperature $T$ by 
adiabatically changing $A_0$ into
$$A\equiv \left[{1\over\beta}
\arcoth\left(2 A_0^{1/2} D_0\right)\right]^2.\eq$$
However, we are not interested in such ``fine-tuned" models,
where $A(t)$ is chosen to depend on the initial data $D_0$.
Rather, we are looking for a fairly robust heat
bath model, where one single interaction potential can thermalize
fairly general initial states.

\header{Initially uncoupled oscillators}

As a simple example, let us choose 
the initial state to be the thermal state at
some temperature $T_0$ for a Hamiltonian given by $A_0=I$. Since
this corresponds to uncoupled oscillators, this state is simply
a direct product of single-oscillator states: $\mu=0$ and 
$C=\oh\coth\beta_0 I$, where $\beta_0\equiv\hootktz$. 
Now let a cyclic interaction potential $A$ be switched on, either
adiabatically or suddenly.
After a while (after the switch-on is complete in the
adiabatic case or after the oscillations have died down in the
sudden case), we can compare the 
resulting steady states with the thermal state:
\CorrEq=\eqnr
$$C = \cases{
C_a = \oh
\pmatrix{A^{-1/2}&0\cr
0&{A^{1/2}}}
\coth\beta_0
&after adiabatic change,\cr
\noalign{\vskip 6pt}
C_s = \oh
\pmatrix{{I+A^{-1}\over 2}&0\cr 0&{I+A\over 2}}
\coth\beta_0
&as $t\to\infty$ after sudden change,\cr
\noalign{\vskip 6pt}
C_t = \oh
\pmatrix{A^{-1/2}\cth&0\cr
0&{A^{1/2}\cth}}
&for a thermal state.\cr
}\eq$$
We see that these three covariance matrices are in general all
different. In particular, the final states in the adiabatic and
sudden cases are not thermal. 
Although there is no reason to expect the total correlation
matrices to be thermal, one might hope that certain
submatrices of $C_a$ or $C_s$ should be thermal, in the
sense that they equal the
corresponding submatrices of $C_t$ --- at least if $C_t$
is allowed to refer to a different temperature, {\ie} allowing
$T\neq T_0$. 
In other words, one might hope that certain subsystems would
approach the state that they would have if the total system
were thermal
\footnote{$^\dagger$}
{This is perhaps the only reasonable definition of what to
mean by a subsystem becoming thermalized.
Recall that 
we have only defined thermality of a system {\it relative to
some Hamiltonian}. Thus if a subsystem in some sense becomes
thermal, the question is: Relative to which Hamiltonian? 
It would be naive to expect the answer to be the
Hamiltonian describing only the forces within the
subsystem, since when a subsystem is coupled to its surrounding,
the Hamiltonian that governs it by definition involves parts
outside of the subsystem.}.
It should be fairly obvious that this is not the case for generic
$A$. 
The nearest-neighbor coupling is a simple counterexample.
For this simple case, we can evaluate the momentum correlations
explicitly in the limit $\gamma\gg 1$, $kT\gg\hbar\omega_0$:
$$\expec{\op_m \op_n} \approx \cases{
\left({\gamma\coth\beta_0\over 2\pi}\right)
{1\over 1 - 4(m-n)^2}&after adiabatic
change,\crr 
\left({\coth\beta_0\over 2}\right)
\left[(1+\gamma^2)\delta_{m,n} -
{\gamma^2\over 2}(\delta_{m,n-1} + \delta_{m,n+1})\right]
&after sudden
change,\crr {1\over 2\beta}\delta_{m,n}&in thermal case.
}\eq$$
(The adiabatic result has been derived using Eq. 
$(\number\FKMeq)$ and
$\gamma\gg 1$, the sudden result is exact and the thermal
result is only valid  for $kT\gg\hbar\omega_0$.)
We see that the qualitative behavior is quite different in all
three cases: 
In the adiabatic case, weak long-range correlations exist. In the
sudden case, correlation exist only between nearest neighbors.
In the high-temperature thermal case, there are no correlations
at all.

For $|m-n|\ll\gamma$, the integral $(\number\FKMeq)$ for position
correlations is completely dominated by the region around
$\theta = \gamma$, and we obtain  
$$\expec{\oq_m \oq_n} \approx \cases{
{\ln(2\pi\gamma)\over 2\pi\gamma}&after adiabatic change,\crr
{1\over 4}\delta_{m,n} + {1\over 8\gamma}
&after sudden
change,\crr 
{1\over 4\gamma\beta}&in thermal case.
}\eq$$
Note that for the adiabatic
and thermal cases, this is independent of $m$ and $n$ , which
physically means that oscillators with
separation  $|m-n|\ll\gamma$ move as one rigid unit, having
their positions almost perfectly correlated.

The simple $2\times 2$ submatrices corresponding to a
single oscillator are all diagonal, just as they are for
the thermal state of a single oscillator. 
Since they contain only two non-zero elements, they are indeed 
identical to the thermal state of a single oscillator with some 
temperature $T$ and coupling $\omega_{eff}\neq\omega$, or 
identical to the $2\times 2$ submatrices of $C_t$
with some temperature $T$ and coupling $\gamma_{eff}\neq\gamma$.
In this limited sense, both adiabatic and sudden change
can be said to thermalize $n=1$ subsystems. 
For subsystems with $n$ oscillators, $C_a$, $C_s$ and $C_t$
each have $n(n+1)$ different non-zero elements in the
submatrix, so we see that the above-mentioned match by
adjusting the two parameters $T$ and $\gamma_{eff}$ was
just a fortuitous coincidence that worked when $n=1$.

Another popular choice of initial conditions in the literature
has been a thermal state with some coupling $A_0\neq I$. Also
this case can be readily treated with the above formalism, and
generally fails to yield thermal states for $n>1$ subsystems.

Does this mean that cyclic quadratic Hamiltonians
are totally incapable of acting as heat baths for $n>1$, or is
there some simple limit in which subsystems can be thermalized?
One frequently discussed limit is that of weak coupling.

\header{The weak coupling limit}

In the case of
infinitesimally weak coupling, say $A = I + \epsilon F$,
Equations $(\number\CorrEq)$ reduce to  
\WeakCorrEq=\eqnr$$C = \cases{
\pmatrix{I-\eot F&0\cr
0&I+\eot F}
{\coth\beta_0\over 2}
&after adiabatic change,\cr
\noalign{\vskip 4pt}
\pmatrix{I-\eot F&0\cr
0&I+\eot F}
{\coth\beta_0\over 2}
&as $t\to\infty$ after sudden change,\cr
\noalign{\vskip 4pt}
\pmatrix{I-\left[1+\sinhmess\right]\eot F&0\cr
0&I+\left[1-\sinhmess\right]\eot F}
{\coth\beta\over 2}
&for a thermal state,\cr
}\eq$$
where 
$$\sinhmess \equiv
{2\beta\over\sinh 2\beta}.\eq$$
Thus we see that for infinitesimally weak coupling, the sudden
case coincides with the adiabatic case, but that 
the resulting state is generally still not thermal.
It should come as no surprise that
the sudden case coincides with the adiabatic
case in this limit, since a sudden infinitesimal change is in a
sense adiabatic.
We see that there are only two cases when the resulting
state is thermal:

The first is if $F$ is a multiple of $I$, which gives 
$T\neq T_0$ and  
corresponds to the trivial case of the oscillators remaining
uncoupled.

The second is when $\sinhmess=0$, which implies $T = T_0 = 0$.
This simply reflects the fact that adiabatic change changes
the ground state into the new ground state. Note that this is a
pure quantum phenomenon, since 
$\sinhmess\to1\neq 0$ as $\hbar\to 0$.

\header{The sense in which harmonic heat baths work}

Although the above results for the weak-coupling limit showed
that subsystems did not become thermal, they did indeed become
almost thermal, the difference being terms of order $\epsilon$.
This is not very impressive at all, in view of the fact that the
initial state was also almost thermal.
However, a general feature of infinite harmonic chains is that
the final state after sudden change remains the same even if the
initial states of a finite number of oscillators is changed.
This follows directly from the fact that all components of
$U(t)$ approach zero as $t\to\infty$. This means that a
sudden change will result in the same almost thermal state even
if some subsystem starts out in a completely different, quite
non-thermal state.  It is in this sense that the many harmonic
heat bath models in the literature can transform a subsystem
into an approximately thermal state.

Unfortunately, the interesting case where the
system has internal couplings stronger than
those of the heat bath cannot be treated with
the formulas derived in this paper, since the matrix $A$ would
not be cyclic. It would appear, however, that the strength of
the coupling between the system and the heat bath must be
carefully balanced for such a heat bath model to work. If the
coupling is so strong that it is comparable to the couplings
within the system, then the final state of the system will
probably deviate considerably from the thermal state. If, on the
other hand, the coupling is too weak, then the relaxation time,
the time scale over which the system approaches the thermal
state, will be unreasonably long.

\beginsection{VIII. THE INDEPENDENCE OF DISPLACEMENT AND
DISPERSION}

In this section, we will discuss how the previous results for
sudden change are modified in the case of
non-zero means, {\ie} when $\vm\neq 0$. It will be seen that the
fact that $\mu$ and $C$ evolve independently of each other makes
systems with $\vm\neq 0$ totally unable to produce thermal
states.

We can generalize $(\number\HamEq)$ to arbitrary quadratic
Hamiltonians by writing
$$H = \oz^TB\oz = \tr B\oz\oz^T,\eq$$
where $B$ is any symmetric matrix. This covers both cases
where the different particles have different masses and cases
where the potential is not positive definite. $$\expec{H} =  \tr
B\expec{\zzt} = \tr B\mmt + \tr BC = E_1 + E_2,\eq$$ say, where
we will call $E_1\equiv \tr B\mmt$ the {\it displacement energy}
and and  $E_2\equiv \tr BC$ the {\it dispersion energy}. 
(To be precise, the second equal sign is valid only if the two
off-diagonal $n\times n$ submatrices of $B$ are symmetric and
thus identical --- otherwise the non-commutativity of $\oq$ and
$\op$ gives rise to an uninteresting extra additive constant.) 
We see from $(\number\EvolEq)$ that $\vm$ and $C$ evolve
completely independently of each other.  By formally setting
either $\mu$ or $C$ equal to zero,  $(\number\EvolEq)$ thus
shows that the displacement energy and the dispersion energy are
{\it separately} conserved, independently of one another. Hence
displacement energy can never be used to increase dispersion and
entropy.  This is in stark contrast to an approach to the
thermal state, where the resulting equilibrium entropy for
subsystems is the maximum allowed by the total available energy.
This difference between the thermal state and the actual state
as $t\to\infty$ is illustrated in Figure 1. The Gaussian Wigner
distribution in the phase-space of a single oscillator is shown
by the contour at which it has dropped to half its maximum
value, a circle. Recall that the entropy of the oscillator
depends on the area of the circle, whereas the energy depends on
the average distance to the origin. Thus the circle symbolizing
the equilibrium state has the maximum area consistent with the
available energy. The actual state has the same energy but much
less entropy.

Thus oscillator systems which are able to evolve a state with
non-zero means into something resembling the thermal state
cannot have purely quadratic Hamiltonians, but must contain some
non-linear couplings.

\vfill\eject

\beginsection{IX. CONCLUDING REMARKS}

We have studied
steady states of coupled harmonic oscillators, and how adiabatic
and sudden changes in a cyclic potential transform one steady
state into another. Our main conclusions are as follows:

\llist{
(i) Adiabatic change can transform a given steady state into any
other steady state, subject to the constraint that the adiabatic
invariant $AD^2$ is conserved.
}

\llist{
(ii) After a sudden change, the covariance matrix undergoes an
oscillatory phase, after which it generally converges to a
new steady state. This asymptotic behavior depends crucially on
how degenerate the spectrum of $A$ is. Generic sudden changes are
not reversible, and a sequence of sudden changes, where the
final and initial potentials $A$ are the same, leads to a net
increase in energy.
}

\llist{
(iii) The states resulting from adiabatic and sudden changes are
in general not thermal states, not even if we restrict our
attention to subsystems.
We conclude that harmonic heat bath models can lead to
approximately thermal subsystems only under the following
conditions:
}

\lllist{
1) The coupling between the system and the heat bath is weak.
}

\lllist{
2) $\mu=0$, {\ie} the expectation values of the displacements and
momenta of the oscillators vanish. }

We summarise the state of affairs for heat bath models as
follows: 
It is frequently assumed that the density matrix of a
system is $\rho\propto e^{-H/kT}$, {\ie} the thermal state
corresponding to a temperature $T$.
Unfortunately, no completely solvable
heat bath model has ever been found that explicitly evolves
multiparticle ($n>1$) systems into such states. 
The problem appears to arise when the coupling
between the system particles and individual heat bath particles
is so strong that it is comparable to the couplings within the
system. 
Thus approximately thermal states can be created only when the 
when the system particles are very weakly coupled to the bath
particles. 
If this is the case, the system particles must be coupled to
very many heat path particles to avoid the approach to
equilibrium taking an unreasonably long time. 
Only in the limit when a typical system particle
becomes infinitesimally coupled to infinitely many
bath particles (such as in the limit of the FKM model
$[\number\FKM]$ or the IO model $[\number\FLO]$, both dealing
with the $n=1$ case) is the limiting state exactly thermal. 
Thus we may expect nature to be full of 
approximately thermal states, but devoid of 
exactly thermal states.

\beginsection X. ACKNOWLEDGMENTS

The authors would like to thank 
Prof. H. Shapiro for many useful suggestions.

\beginsection{APPENDIX A}

In this appendix, we will discuss the circumstances under
which  $\Cosc(t)\to 0$ as $t\to\infty$.

Applying Eq. $(\number\fmnEq)$ to the components of the various
terms in $\Cosc(t)$, the function $f(\lambda^2)$ always contains
a trigonometric factor that oscillates increasingly rapidly with
respect to $\lambda^2$ as $t\to \infty$. In Section VI, we gave a
qualitative discussion of how this causes $\Cosc(t)$ to approach
zero for general symmetric positive definite $A$ with
non-degenerate spectra, as well as what happens when $N$ is
large but finite. Here we will give a rigorous
treatment for the special case when $A$ is
infinite-dimensional and cyclic. 

The behavior of $\Cosc(t)$ depends crucially on how
degenerate the spectrum of $A$ is. Since all cyclic
Hamiltonians are invariant under parity, they all have the
degeneracy
$$\omega(-\theta) = \omega(\theta),\eq$$
where we have defined the frequency $\omega\equiv |\lambda|$.
If this is the {\it only} degeneracy, {\ie} if 
$\omega(\theta)$ is invertible for $\theta\geq 0$, and if in
addition $\omega(\theta)$ is differentiable, then we can change
variables in Eq. $(\number\FKMeq)$ and obtain
\fmnEqTwo=\eqnr
$$f(A)_{mn} = 
2 \iww f(\omega^2) g(\omega) \cos\left[(m-n)\theta(\omega)\right]
d\omega.\eq$$ 
Here we have introduced the {\it spectral density function}
$g(w)$, defined so that 
$g(\omega)d\omega$ is the fraction of the frequencies
that lie in the interval $[\omega,\omega+d\omega]$.
Note that 
$$g(\omega) = {1\over 2\pi}{d\theta\over d\omega}\eq$$
if $\theta(\omega)$ is single-valued and invertible, but that
$g(\omega)$ is well-defined as a distribution for {\it any}
spectrum $\omega(\theta)$. For the case of nearest-neighbor
coupling, 
$$g(\omega) = 
\cases{{1\over\pi} 
{\omega\over\sqrt{\left(\omega^2-1\right)
\left(4\gamma^2 + 1 - \omega^2\right)}}
&if $1\le|\omega|\le 1+4\gamma^2$,\cr
0&otherwise.}\eq$$
We wish to examine the various terms in
$\Cosc(t)$. Let us define 
$$a_n \equiv f(A)_{m,m+n}\eq$$
and choose the function to be
$$f(A) = \cos(A^{1/2}t).\eq$$
Eq. $(\number\fmnEqTwo)$ now yields
$$a_n(t) = 
2 \iww \cos(\omega t) g(\omega) \cos\left[n\theta(\omega)\right]
d\omega.\eq$$
By extending the functions $g$ and $\theta$ to the entire real
line by $g(-\omega)\equiv g(\omega)$,
$\theta(-\omega)\equiv \theta(\omega)$, 
and $g(\omega) = 0$ unless 
$\omega_{min}\le|\omega|\le\omega_{max}$, we see that
this can be written as a
Fourier transform: 
$$a_n(t) = \uh_n(t),\eq$$
where
$$u_n(\omega) \equiv g(\omega)
\cos\left[n\theta(\omega)\right].\eq$$ According to
Riemann-Lebesgue's Lemma,  $$\uh_n(t) \to 0\quad\hbox{as}\quad
|t|\to\infty\eq$$ if $u_n(\omega)$ is an integrable function.
Since the latter is trivially satisfied for all $n$
(indeed $\imii u_n(\omega) d\omega = 2\pi\delta_{0n}$), this
implies that
\footnote{$^\dagger$}
{The convergence is not necessarily uniform, {\ie} although
$\cos(A^{1/2}t)_{mn}\to 0$ for any fixed $m$ and $n$, we do not
always have $\sup_{m,n} \cos(A^{1/2}t)_{mn} \to 0$. 
A more careful analysis
shows that uniform convergence is obtained if the spectral
function $\lambda$ is analytic on the entire interval
$[-\pi,\pi]$ and in addition is nonlinear.}
$$\cos(A^{1/2}t) \to 0\quad\hbox{as}\quad
t\to\infty.\eq$$ 
In an almost identical fashion it is readily shown
that $\sin(A^{1/2}t)\to 0$ as $t\to\infty$. By multiplying
these results by various constant matrices, it follows that all
terms in $\Cosc(t)$ approach zero. In summary, we have shown that
all components of $\Cosc(t)\to 0$ as $t\to\infty$
for any symmetric, positive definite, cyclic,
infinite-dimensional $A$ if its spectral function
$\lambda^2(\theta)$ is invertible and differentiable for 
$\theta > 0$.

It is straightforward to generalize this result in a number of
ways:

\slist{
* If $\lambda^2(\theta)$ is differentiable but not invertible
because it turns around a finite number of times, then the same
result is obtained by integrating over each monotonic segment
separately.}

\slist{
* If $A$ is merely non-negative definite, then
$A^{-1/2}$ and
$A^{-1}$ are undefined. However, the result still holds for the
momentum-momentum part of the covariance matrix.}

Roughly speaking, the key result $\Cosc(t)\to 0$ as $t\to\infty$ 
fails if the spectrum is too degenerate. An easy way to
appreciate this is to study the diagonal terms of the matrix
$\cos(A^{1/2}t)$, which simply equal
$$a_0(t) = \gh(t).\eq$$
For uncoupled oscillators, we have $A\propto I$ and
the spectrum is totally degenerate, {\ie} the
spectral density 
$g(\omega)=\delta(\omega-\omega_0)+\delta(\omega+\omega_0)$ 
and 
$a_0(t) \propto \cos\omega_0 t$ for all $t$.
By Riemann-Lebesgue's Lemma, ${a_0(t)\to 0}$ as ${t\to\infty}$
if the spectral density is an integrable function, not
if it is merely a tempered distribution such as $\delta$.
Furthermore, from partial integration of the Fourier integral,
we know that if  the spectral density is a $k$ times
differentiable function, then $a_n(t)$ approaches zero faster
than  $t^{-k}$.

\vfill
\break
\baselineskip12pt
\frenchspacing

\beginsection  REFERENCES
 
% \narrower
\smallskip
\refnr=1

% HEAT BATH REFS:

\rf M. A. Huerta and H. S. Robertson; J. Stat. Phys.;1;393;1969

\rf E. Schr\"odinger; Ann. Phys. (Leipzig);44;916;1914

\rf G. Klein and I. Prigogine; Physica;19;1053;1953

\refn P. C. Hemmer, {\it Dynamics and Stochastic Types of Motion
in the Linear Chain}, Thesis, University of Trondheim, Norway
(1959).

\rf G. W. Ford, M. Kac, and P. Mazur;J. Math. Phys.;6;504;1965
 
\rf P. Mazur and E. Montroll; J. Math. Phys;1;70;1960

\rf A. O. Caldeira and A. J. Leggett; Ann. 
Phys. (N.Y.);149;374;1983

\rf P. Ullersma; Physica;32;27;1966

\rf A. O. Caldeira and A. J. Leggett; Physica;A121;587;1983

\rf F. Haake and R. Reibold;Phys. Rev.;A32;2462;1985

\rf G. W. Ford, J. T. Lewis and R. F. O'Connell; J.
Stat. Phys.;53;439;1988

\rf G. W. Ford, J. T. Lewis and R. F. O'Connell; Phys.
Rev.;A37;4419;1988

% WIGNER REFS:

\rf E. P. Wigner;Phys. Rev.;40;749;1932

\rf M. Hillery, R. H. O'Connell, M. O. Scully, and
E. P . Wigner;Phys. Rep.;106;121;1984
 
\refn Y. S. Kim and M. E. Noz, {\it Phase Space Picture of
Quantum Mechanics: Group Theoretical Approach} (World
Scientific, Singapore, 1991)

% ADIABATIC REF:

\refn E. J. Saletan and A. H. Cromer,
{\it Theoretical Mechanics} (Wiley, New York, 1971)

\end